\documentclass[journal=langd5,manuscript=article]{achemso}
\AtBeginDocument{\singlespacing}
\usepackage{graphicx}% Include figure files
\usepackage{dcolumn}% Align table columns on decimal point
\usepackage{bm}% bold math

\usepackage[utf8]{inputenc}
\usepackage[T1]{fontenc}
\usepackage{mathptmx}
\usepackage{etoolbox}
\usepackage{siunitx}
\usepackage{subcaption}
\usepackage{PPtools}
\usepackage{import}
\usepackage{chemformula}
\usepackage[version=3]{mhchem}

\setkeys{acs}{doi = true}

\makeatother

\title{Azimuthal variation of apparent contact angles on structured surfaces featuring micrometric ramps, pyramids and staggered cubes at two different inherent wettabilities}
\author{P. Palmetshofer}
\email{patrick.palmetshofer@itlr.uni-stuttgart.de}
\affiliation{ 
Institute of Aerospace Thermodynamics, University of Stuttgart, Pfaffenwaldring 31, 70569 Stuttgart, Germany}%
\author{S. Hengsbach}%
\author{M. Guttmann}%
\author{M. Worgull}%
\affiliation{Institute of Microstructure Technology (IMT), Karlsruhe Institute of Technology,
76344 Eggenstein-Leopoldshafen, Germany}
\author{B. Weigand}%
\affiliation{ 
Institute of Aerospace Thermodynamics, University of Stuttgart, 70569, Germany}%

\date{\today}

\begin{document}
\maketitle
\begin{abstract}
As new manufacturing methods enable manufacturing of microstructured surfaces with varying structure geometries, questions remain on some effects on the wetting behavior and the resulting apparent contact angles.
In this study, we report the manufacturing process using 3D Direct Laser Writing (3D-DLW) and hot embossing for Poly-methylmethacrylate (PMMA) surfaces with micrometric pyramids, cubes on a staggered grid and two types of ramped structures.
We measure the azimuthal variation of the apparent contact angle of sessile droplets on the surfaces. Using plasma polymerization or no treatment of the surfaces, two different inherent wettabilities are studied.
We find that while all structure types cause an azimuthal variation of the apparent contact angle, pinning at the ramp tops increases the contact angle more strongly on one side.
On pyramid structures, pinning lines can occur on the structure axes and diagonals similarly. For cubes on a hexagonal grid, the strongest contact angle increases are observed along the primary structure axes, where pinning is preferred while smaller peaks are seen on the secondary axes at $60^{\circ}$ and $120^{\circ}$ to the primary axis.
\end{abstract}

\section{\label{sec:intro}Introduction}
The wettability of a surface is an important parameter in a variety of applications, such as spray cooling, spray painting or spray coating, making it a focus in surface engineering research\cite{Quere2005}. The wettability of a surface can intentionally be modified by varying the surface material or chemistry or varying the surface roughness\cite{Huang2020,Kalinin2009,Gennes1985}. Similarly to surface roughness, microstructures can be introduced on the surface, which can alter the wettability even anisotropically\cite{Foltyn2021a,Courbin2007}.
Most research on the effects of microstructures on sessile and dynamic droplets, however, has either considered isotropic roughness\cite{Massinon2017,Quere2008} or simple structure elements such as cylinders\cite{Li2013}, cuboids\cite{Foltyn2021b} or grooves\cite{Jin2020}. Asymmetric surfaces such as inclined pillars\cite{Yada2021,Yada2022,Lee2019} or differently shaped surface structures such as ramps and pyramids have been studied only rarely. As such, there is a notable gap in the understanding of how such structures can affect wetting and imbibition into such structures.

On ideally smooth surfaces, the surface chemistry determines the contact angle of a droplet, resulting in a spherical cap-shaped droplet if the droplet is small enough\cite{Shafrin1960}. The resulting equilibrium contact angle is also called Young's contact angle $\theta_Y$. This contact angle directly follows from the balance of surface energies acting on the contact line.

\begin{figure}
    \centering
    \includegraphics{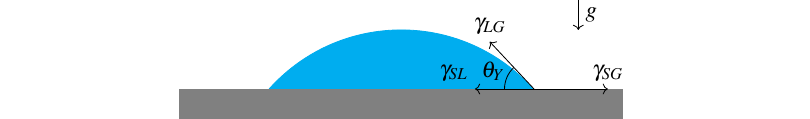}
    \caption{Definition of Young's static contact angle $\theta_Y$ on a smooth surface using the liquid-gas $\gamma_{LG}$, solid-liquid $\gamma_{SL}$ and solid-gas $\gamma_{SG}$ surface energies.}
    \label{fig:YoungCA}
\end{figure}
As illustrated in figure~\ref{fig:YoungCA}, they depend on each other through the relation
\begin{equation}
   \gamma_{LG} \cos \theta_Y = \gamma_{SG} - \gamma_{SL}
\end{equation}
where $\gamma_{LG}$ is the liquid-gas surface energy (the surface tension), $\gamma_{SG}$ is the solid-gas surface energy and $\gamma_{SL}$ is the solid-liquid surface energy. In the scope of this study, the measured apparent contact angles on flat portions of the surface $\theta_{FS}$ are treated as an estimate for Young's contact angle $\theta_Y$.
Using similar energetic arguments for an equilibrium state, two different macroscopic apparent contact angles can be derived for two different states. In the Wenzel state, which is defined as the liquid wetting the complete surface area, the apparent contact angle $\theta_W$ in the Wenzel state\cite{Wenzel1936} is modified as
\begin{equation} \label{eq:Wenzel}
\cos \theta_W = r \cos \theta_Y
\end{equation}
through a roughness factor $r=A_{total}/A_{proj}$, which is the total (rough) surface area $A_{total}$ divided by the projected (equivalent flat) surface area $A_{proj}$.
In the Wenzel state, the apparent contact angle is smaller than the equilibrium contact angle $\theta_W < \theta_Y$ for $\theta_Y < 90^{\circ}$ and larger than it $\theta_W > \theta_Y$ for $\theta_Y < 90^{\circ}$ as it amplifies the inherent wettability of a surface.
Whether the Wenzel state is stable depends on the structure and inherent wettability. Bico et al.\cite{Bico2002} define a critical contact angle for transition from the Wenzel state to a Cassie-Baxter state as
\begin{equation} \label{eq:WCBtrans}
    \cos \theta_{Y,crit} = \frac{\Phi_S-1}{r-\Phi_S}
\end{equation}
where $\Phi_S=A_{top}/A_{proj}$ is the area of the pillar tops divided by the projected surface area. In the Cassie-Baxter state, the droplet only wets the pillar tops and the gaps between pillars are filled with air. In this state, the droplet assumes an apparent contact angle of\cite{Cassie1944}
\begin{equation}\label{eq:CB}
    \cos{\theta_{CB}} = \Phi_S (1+\cos{\theta_Y})-1,
\end{equation}

On regularly structured surfaces, additional variations of the apparent contact angle of a sessile droplet can occur due to pinning at structure elements\cite{Foltyn2021b}. To characterize the effect of the surface structures on a static contact angle, Foltyn et al.~\cite{Foltyn2021} developed a method to measure the contact angle at varying azimuthal angles over the full range between $0^{\circ}$ and $360^{\circ}$. 
In the past, numerous authors used either grooves, cuboid structures or cylindrical structures on square grids \cite{Foltyn2019b,Broom2022,BroomPhD,Li2013}. This includes Foltyn \cite{FoltynPhD}, who used cuboid structures with a height of $h=\qty{20}{\micro\meter}$ and side lengths of \qty{60}{\micro\meter}, \qty{30}{\micro\meter} and \qty{15}{\micro\meter} in addition to grooves with a width and separation of \qty{60}{\micro\meter} and a length of \qty{500}{\micro\meter}. Along the structure axes and at the groove sides, an increase in the apparent contact angle was observed originating from pinning at the pillar tops. When plotting the apparent contact angles in a polar plot, a rhomboid-like shape was observed for the cuboids, with the sides at $45^{\circ}$ to the structure axis.
For the grooves, larger contact angles are observed at the sides.
In regions of larger contact angles, a shorter contact line was observed, which follows from the tendency of the droplet to form a spherical-cap like shape at its top. A shorter contact line can thus only be reached by larger curvature in its vicinity, leading to a larger contact angle. 

As this study aims to study the effect of the structure shape, comparable structure sizes as those used by Foltyn et al.\cite{Foltyn2021b} are chosen.
In this work, we study ramps and sharp-tipped pyramids, structures which do not feature a flat pillar top that allows symmetric pinning at this area. Additionally, we use micrometric cubic pillars in a staggered arrangement, corresponding to a hexagonal grid.

In the following, the surface structures, their manufacturing process and characterization is described in detail. Subsequently, the measurement method for the azimuthal variation of the contact angle is given and both the contact angle variations and contact lines are given and discussed.

%As discussed in the study of Foltyn et al.~\cite{Foltyn2022} on dynamic droplet impacts, we also use plasma-treated surfaces. As no apparent contact angle is measurable for plasma-activated surfaces, only plasma polymerization is used, which increases Young's contact angle to $\theta_{FS}= 120^{\circ}$

\section{\label{sec:meth}Experimental methods}
\subsection{Investigated surface types}\label{sec:types}
%The surfaces were manufactured by the Karlsruhe Nano Micro Facility using hot embossing of poly-methylmethacrylate (PMMA) with a silicon stamp manufactured through 3D-Direct Laser Writing (3D-DLW).
%The microstructures were developed and manufactured by the \textit{Karlsruhe Nano Micro Facility (KNMF)}, who also analyzed the structures.

\begin{figure}
    \centering
    \includegraphics{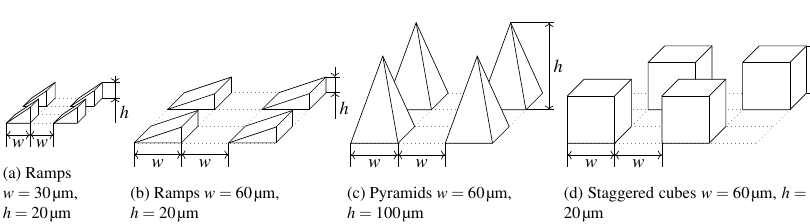}
    \caption{Investigated surface structure shapes and sizes and backlight images of the structures}
    \label{fig:newstructures}
\end{figure}

\begin{table}[h]
\small
\centering
	%PMMA surface types:\\
	\begin{tabular}{|l|c|c|c|}
		\hline 
		Surface Type		          & $r$ 		& $w/h$ & $R_z / \unit{\micro\meter}$ \\ \hline 
		Ramps \qty{30}{\micro\meter}   & 1.38 & 1.5 & 20 \\ \hline
		Ramps \qty{60}{\micro\meter}   & 1.18 & 3.0 & 20 \\ \hline
		Pyramids				      & 1.62 & 0.6 & 100 \\ \hline
		Staggered Cubes       	      & 2.00 & 1.0 & 60 \\ \hline
	\end{tabular}
	\caption{Geometric roughness parameters of the surfaces}
	\label{tab:GeoParametersMicro}
\end{table}
The four investigated structures with relevant lengths are shown in figure~\ref{fig:newstructures}. The ramps have a feature height of $h=\qty{20}{\micro\meter}$, making them comparable to either the \qty{60}{\micro\meter} or \qty{30}{\micro\meter} cuboids used by Foltyn~\cite{FoltynPhD}. The cubes have a side length, height and separation of \qty{60}{\micro\meter}. The separation and side length of the pyramids is also \qty{60}{\micro\meter}, but their height of \qty{100}{\micro\meter} makes it possible to discuss some effects of the aspect ratio. The cubes used here are placed in a staggered arangement, meaning that each row is displaced by half the repeating feature length. In other words, this means the cubes are placed on a hexagonal grid, while all other structures are arranged on a square grid.

% \begin{figure}
%     \centering
%     \input{content/Figures/Methods/Images_Microstructures/Images_Microstructures}
%     \caption{Backlight images of the structures, courtesy of KNMF}
%     \label{fig:newstructuresimages}
% \end{figure}
Table~\ref{tab:GeoParametersMicro} provides an overview of the roughness factors $r$ for the given surface types, the aspect ratio of the structures $w/h$ and the ten-point-height roughness $R_z$, which in this case is equal to the maximum vertical distance between the highest points and lowest points of the structure. Note that the \qty{60}{\micro\meter} ramps feature the lowest roughness factor due to their high aspect ratio.

\subsection{Surface manufacturing}

% --------------------------------------
%\subsection{Fabrication}

\begin{figure}
    \centering
     %\begin{tikzpicture}
    %\node at (0,0) {\includegraphics[width=0.4\linewidth]{Figures/fabricationprocess.png}};
	%gravity
    %	\draw [->,draw=black] (3,2) -- (3,1.5) node [midway,right] {$g$};
    % \end{tikzpicture}
\includegraphics{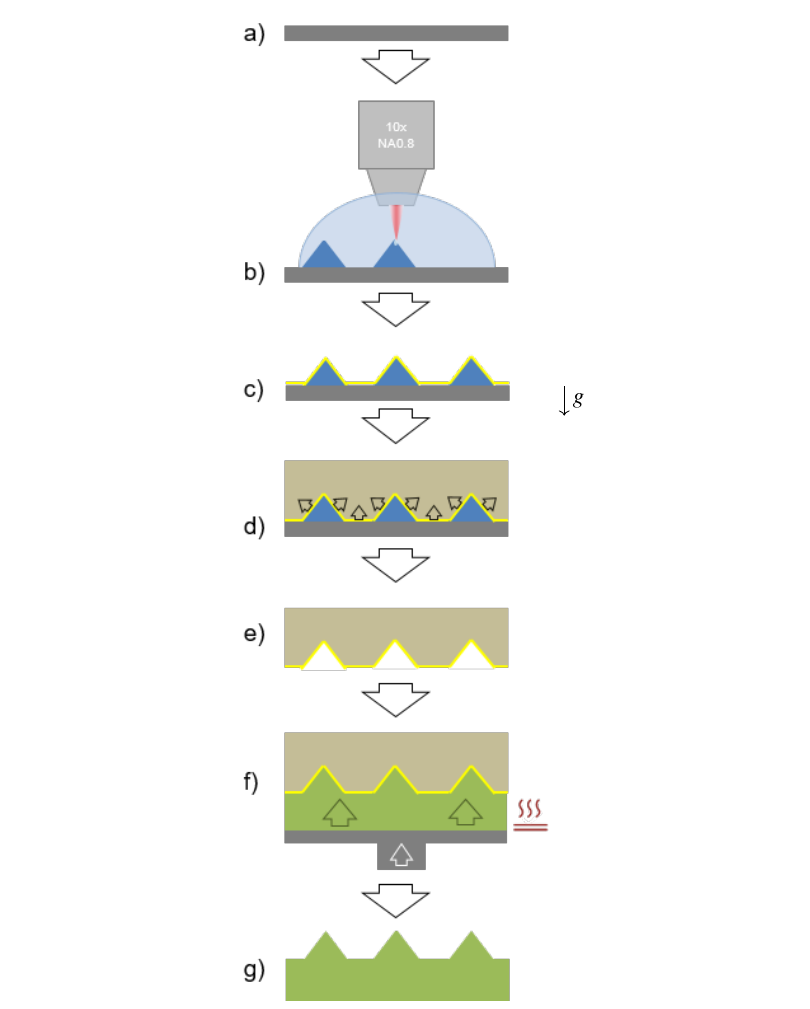}
    \caption{Process chain of fabrication; a) 4“ Si-Wafer with silanized surface; b) Structuring by 2-Photon-Polymerization c) Application of conductive layer for electroplating d) Electroplating of nickel e) Desolving wafer with Potassium hydroxide (\ce{KOH}) and stripping resist with \ce{O2}-Plasma; f) Replication in PMMA by hot embossing g) Replicated functional polymer structures on a carrier layer}
    \label{fig:FabProcess}
\end{figure}

\begin{figure}
    \centering
    \includegraphics{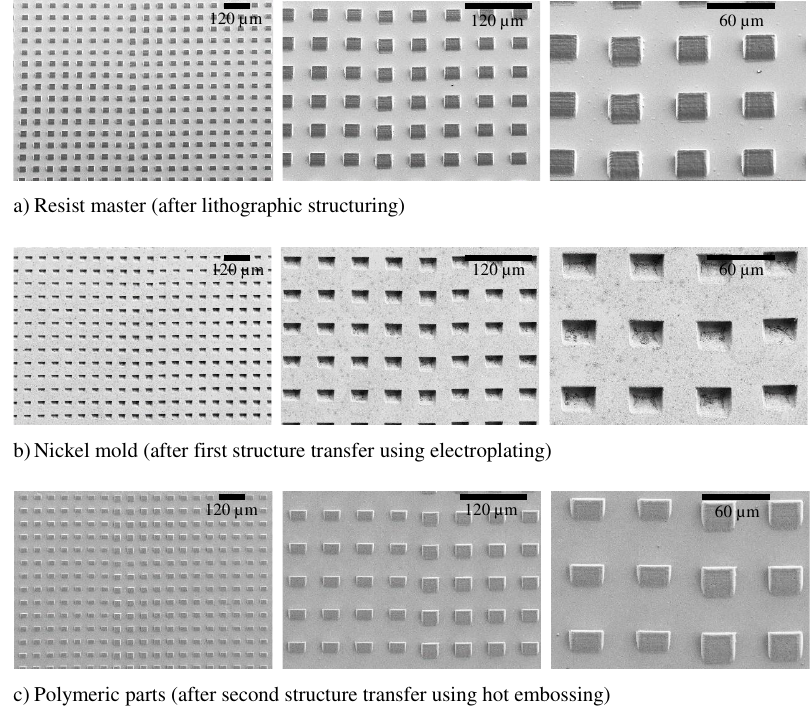}
    \caption{SEM pictures of the \qty{30}{\micro\meter} ramps throughout the manufacturing process chain a) Master b) Nickel mold c) replicated PMMA parts}
    \label{fig:SEM-structures-rect}
\end{figure}

\begin{figure}
    \centering
    \includegraphics{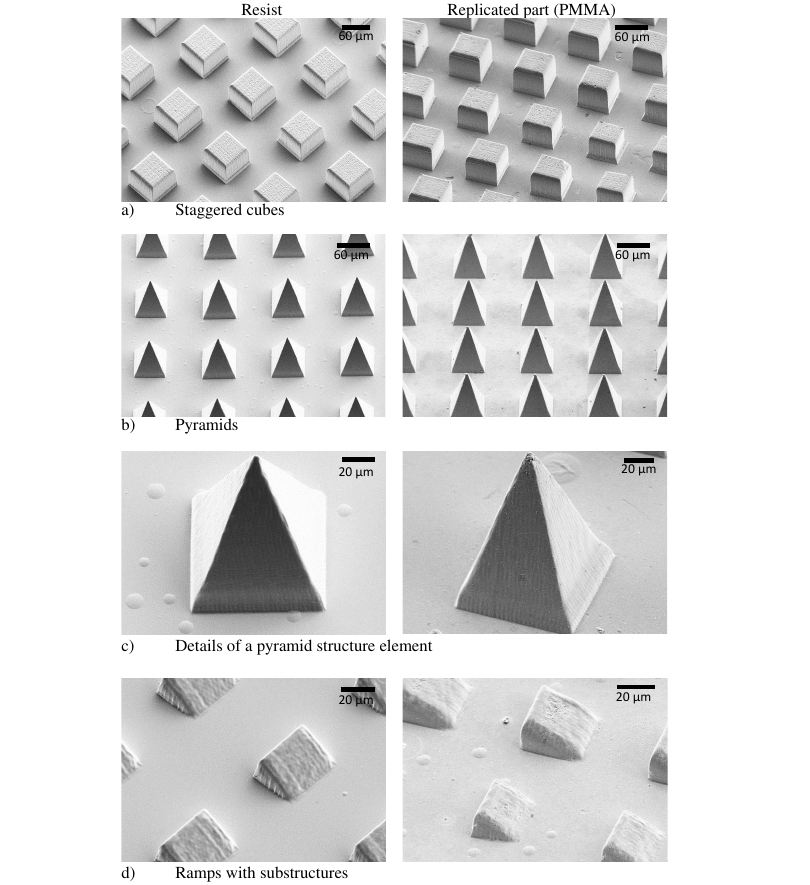}
    \caption{SEM pictures of all structures. The left column shows the structured resist, the right column illustrates the replicated PMMA samples. All details of the master structures were transferred by electroplating and hot embossing. a) staggered cubes ($w=h=\qty{60}{\micro\meter}$), b) pyramids height ($w=\qty{60}{\micro\meter}$, $h=\qty{100}{\micro\meter}$) c) details of pyramids d) $w=\qty{60}{\micro\meter}$ ramps ($h=\qty{20}{\micro\meter}$; the double-structure defects shown here are at the edges of the structured region and are avoided during droplet deposition). The substructures from the master process were also transferred through the full process chain from electroplating to hot embossing.}
    \label{fig:SEM-structures-pyramid}
\end{figure}

To manufacture the surfaces, 3D Direct Laser Writing (3D-DLW) based on 2-Photon-Polymerization (2PP) was used to apply structures to a silicon wafer. A negative was then manufactured through Nickel Electroplating, which served as a stamp for a hot embossing step. The complete process chain is illustrated in figure \ref{fig:FabProcess}. To ensure the quality of all process steps, scanning electron microscopy (SEM) images were captured for the structured wafer, the nickel mold and the final PMMA samples. They are depicted in figure \ref{fig:SEM-structures-rect}- \ref{fig:SEM-structures-pyramid} for all four surface types. 

% ------------------------
\subsubsection{Preparation}

To avoid deformation during the electroplating process, a thickness of \qty{2}{\milli\metre} was selected for the 4" silicon wafers. The surfaces were pre-cleaned using an oxygen plasma (4Tec) at \qty{100}{\watt} and a flow rate of ~55 SCCM to remove impurities from the surface. This also removes moisture from the surface which can otherwise cause adhesion issues. To further improve adhesion, the substrate was silanized by treating the wafer for \qty{2}{\minute} in a solution of Trimethoxysilylpropyl methacrylate. Afterwards the substrate was rinsed gently with pure Ethanol and tempered for \qty{3}{\minute} on a hot plate at \qty{110}{\degreeCelsius}. Subsequently, the liquid photoresist was applied to the substrate by covering the area to be structured. 

% --------------------------
\subsubsection{Structuring}
%%TODO
To structure the wafer, a Nanoscribe GmbH Photonic Professional GT2 Plus was used in the Large Feature Set configuration (LF-Set). This contains a 10 x NA0.8 Objective in combination with IP-Q Photoresin (Nanoscribe GmbH). The objective is dipped in the liquid photoresin and due to the principle of 2-Photon absorption the resin is only exposed at the center of the focal point. In this setup, resolutions of \qty{1}{\micro\metre} can be achieved. By scanning the beam in the field of view (FoV) of the objective in combination with laterally moving the substrate with stages structures can be printed. 
To print the structures, stereolitography (STL) files are loaded into the software \textit{Describe} provided with the printer. The software separates the design in slices, followed by a separation in hatches on each layer. For the fabrication an adaptive slicing between \qty{0.5}{\micro\metre} and \qty{5}{\micro\metre} and a hatching of \qty{1}{\micro\metre} in combination with an additional contour line was selected to strike a balance between quality of the structures and printing time. All structures were printed onto a $\approx$ \qty{20}{\milli\metre} x \qty{20}{\milli\metre} field. Since this size exceeds the FoV of the objective the field was separated in 20 x 20 fields with a side length of \qty{960}{\micro\metre} each. To make the best use of the available space on the 4" Wafer, the four structure types were printed onto the same wafer with a separation of \qty{10}{\milli\metre}.
An optical laser power of \qty{25.5}{\milli\watt} and a writing speed of \qty{10}{\milli\metre\per\second} was used, resulting in a fabrication time  of 90 hours per field.
%Diskussion an dieser Stelle  warum die fertigung einer so großen Fläche aufgrund der langen herstellungszeit problematisch ist? -> Stabilität, blasenbildung im resist-> Druckabbruch? Ist u.U. für den Leser nicht interessant(?).

% --------------------------
\subsubsection{Development}

%25min PGMEA neu + 25min PGMEA neu + 2min Novec + Kühlschrank
After the fabrication of the structures the unpolymerized resist was removed. For this, the substrate is submerged in 1-Methoxypropan-2-yl acetate (PGMEA) for \qty{25}{\minute}. This step is repeated once with unsaturated PGMEA. Subsequently, the substrate is dipped into NOVEC 3100 (3M) for \qty{2}{\minute} and then dried in a fridge at \qty{7}{\degreeCelsius} for \qty{15}{\minute}. This step reduces delamination of the structures from  the substrates due to high capillary forces while drying of the solvent. NOVEC is an engineered fluid with minimal surface tension which reduces these forces drastically.

% --------------------------
\subsubsection{Metallization, electroplating and final treatment}

At first, chromium (\qty{7}{\nano\metre}) and gold (\qty{50}{\nano\metre}) layers were evaporated onto the wafer and the polymeric resist structures. The chromium layer serves as an adhesive layer while the gold layer acts as a conductive plating base for the subsequent deposition of nickel. During metallization using the PVD process, the substrate was rotated and tilted at \ang{30}. 
For the galvanic nickel deposition, the metallized microstructured wafer was fixed on a special wafer holder and immersed in an electrolytic bath. Nickel electroplating was carried out in boric acid containing sulphamate electrolyte (pH 3.4 to 3.6 at \qty{52}{\degreeCelsius}) for $\approx 287$ hours. To ensure a slow growth of the nickel layer and to achieve a defect-free filling of the lithographic generated microstructured areas, the current density was adjusted to \qty{0.25}{\ampere\per\deci\metre\squared} (corresponding a growth speed of approximately \qty{0.05}{\micro\metre\per\minute}) at the beginning of the plating process. After \qty{60}{\minute}, the current density was increased to \qty{0.5}{\ampere\per\deci\metre\squared} (approx. \qty{0.1}{\micro\metre\per\minute}) and in further steps up to \qty{1.25}{\ampere\per\deci\metre\squared} ($\approx \qty{0.25}{\micro\metre\per\minute}$). Once a nickel height of approximately \qty{3000}{\micro\metre} was reached, the galvanic process was completed and the composite of thick silicon substrate and thick nickel layer was removed from the holder. In a next step, the final height of the nickel layer (and thus the subsequent mold insert) was set to \qty{2.0}{\milli\metre} using wire-cut electrical discharge machining (EDM).
Subsequently, the silicon substrate was separated by wet chemical etching in a \ce{KOH} solution (30 wt.-\% in water; \qty{80}{\degreeCelsius}) combined with a simple lift-off process. The resist, contained in the cavities of the nickel mold, was removed by plasma treatment (oxygen plasma, \qty{60}{\minute}, \qty{60}{\degreeCelsius}, \qty{1200}{\watt}). Metallization layers (gold and chromium) were not etched and remain on the surface of the mold. The final product is a stiff and homogenous metal plate, able to withstand the forces occurring during the hot embossing process.

% --------------------------
\subsubsection{Hot embossing}
Due to the thin carrier layer of the structures, the replication of the electroplated shim was performed by hot embossing, well suited for large area replication on thin polymer layers. For this, the shim was aligned and mounted inside an embedded tool of the hot embossing machine HEX03 (Jenoptik Microtechnik) in the opposite of the substrate plate, which is a sandblasted steel plate. Between the shim and the substrate plate a PMMA polymer film with an area of \qty{60}{\milli\metre} x \qty{60}{\milli\metre} and a thickness of \qty{1}{\milli\metre} was aligned. The stack of shim, polymer film and substrate plate was clamped with a touch force of $\approx \qty{100}{\newton}$ to achieve an efficient heat transfer from the heating units into the stack. After the embossing temperature of \qty{165}{\degreeCelsius} was achieved the embossing was performed in two steps: a velocity controlled (around \qty{1}{\milli\metre\per\second}) step until a force of \qty{60}{\kilo\newton} was reached. This force was maintained in the second step (dwell pressure) while the stacked cooled down to the  demolding temperature of \qty{75}{\degreeCelsius}. Demolding was performed by a precise relative motion of shim and substrate plate. Finally, with the adhesion of the carrier layer (the residual layer) on the sandblasted substrate plate, the structures are demolded precisely in the direction of the structure orientation without any peeling. However, a transparent residual layer requires e.g. a polished steel plate as substrate plate. Here, a manual demolding can be required due to the loss of adhesion. The process chain depicted in figure~\ref{fig:FabProcess} demonstrates the detailed structure transfer from the lithographic master via electroplating to the replication by hot embossing.

%\subsection{Evaluation of surfaces}
% The surface structures were evaluated using scanning electron microscopy of the final PMMA samples.
% \begin{figure}
%     \centering
%     \scalebarnw[black]{\includegraphics[width=0.4\linewidth]{Figures/Images_Microstructures/SEM_Images/S_21_030550_005_A247.jpg}}{1024}{0.22727}{30}{\unit{\micro\meter}}
%     \includegraphics[width=0.4\linewidth]{Figures/Images_Microstructures/SEM_Images/S_21_030550_005_A240.jpg}
%     \includegraphics[width=0.4\linewidth]{Figures/Images_Microstructures/SEM_Images/S_21_030550_005_A244.jpg}
%     \includegraphics[width=0.4\linewidth]{Figures/Images_Microstructures/SEM_Images/S_21_030550_005_A251.jpg}
%     \caption{SEM images of the PMMA surface structures manufactured through 2 Photon Polymerization (2PP) of a silicon wafer and hot embossing}
%     \label{fig:SEMsurface}
% \end{figure}
% Figure~\ref{fig:SEMsurface} shows representative scanning electron microscopy (SEM) images of the surface samples. Overall a good reproduction quality is achieved, however, especially for the smaller \qty{30}{\micro\meter} ramps some variations in the final structure element size are visible. Additionally, for some manufactured surface samples the edges show torn off cubes for the staggered cubic structures. Later, the sessile droplets are intentionally placed in regions where a good reproduction quality was achieved.
\subsection{Plasma polymerization}
For plasma polymerization, a \textit{Diener Electronic Femto} plasma machine with perfluoro monomer was used, which creates a nanometric PTFE-like layer on the surface. The polymerization occurs at a chamber pressure of \qty{10}{\Pa} at a monomer flow rate of \qty{10}{\micro \liter \per \minute}. The generator is pulsed with a duty cycle of 60\% and operates at \qty{13.56}{\MHz} for a duration of \qty{45}{minutes}. The coating applied during plasma polymerization is only several nanometers thick\cite{FoltynPhD} and thus does not significantly affect the surface roughness. However, it fully governs the surface chemical properties, increasing the contact angle on a flat part of the surface to $\theta_{FS}=120^{\circ} \pm 5^{\circ}$ for various underlying surface materials, including PMMA\cite{FoltynPhD}.

\subsection{Contact angle measurement}
%\subsection{Contact angle measurements on flat surfaces} 

\begin{figure}
    \centering   
    %\begin{tikzpicture}
    %\node at (0,0) {\import{Figures}{SchematicCA}};
	%gravity
   % 	\draw [->,draw=black] (5,2) -- (5,1.5) node [midway,right] {$g$};
     %\end{tikzpicture}
 \includegraphics{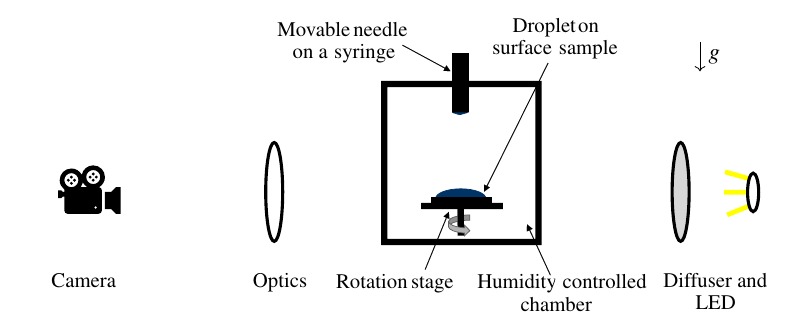}
    \caption{Schematic of the experimental setup for the $360^{\circ}$ contact angle measurement as previously used by Foltyn et al.\cite{Foltyn2021b}. The camera captures side-view diffuse backlight images of the sessile droplet, which is contained in a transparent, humidity controlled chamber. The surface with the sessile droplet is rotated in $1^{\circ}$ steps to obtain images of the sessile droplets from the full azimuthal range.}
    \label{fig:expschematic}
\end{figure}
For the contact angle measurements, the same setup as in the work of Foltyn et al.~\cite{Foltyn2021b} is used, which is shown schematically in figure~\ref{fig:expschematic}. It features an \textit{IDS Imaging Development Systems GmbH UI-3360CP-M} camera with a \textit{Navitar Zoom 6000} lens as part of the system by \textit{DataPhysics Instruments GmbH OCA 15EC}.
%\subsection{360° static contact angle measurements}
On microstructured surfaces, pinning of the contact line on the tops or sides of the structure elements can cause non-circular contact lines even for simple sessile droplets. Similarly, the local contact angle varies along the contact line \cite{Foltyn2022,Palmetshofer2023_dipsi}.
While measuring the real local contact angle is difficult, an apparent contact angle can be measured as a function of the azimuthal angle $\theta_{app}=f(\phi)$. 
For this, the previously described contact angle measurement setup was used together with an electronically actuated rotation stage as in the work of Foltyn et al.~\cite{Foltyn2021b}. 
Additionally, to keep the ambient conditions constant and to avoid evaporation of the droplet during the measurement, the measurement platform in the contact angle measurement device is housed in an enclosed space with transparent windows on each side. Two ultrasound atomizers are controlled by a humidity sensor to keep the relative humidity at \qty{98}{\percent}, preventing evaporative mass loss during the measurement~\cite{Foltyn2021b}. 

% \begin{figure}
%     \centering
%     \includegraphics[width=0.2\linewidth]{Figures/Images_Microstructures/Side_View_NA_R30/Strukt_KNMF_gen3_PMMA_01_R30-333.jpg}
%     \caption{Side view of droplet on ramps visualizing that the contact angle on the left side is smaller than on the right}
%     \label{fig:rampasymmetry}
% \end{figure}

The image analysis was performed using the \textit{DataPhysics SCA20} program. In it, a baseline of the droplet was determined manually and the surface contour was fitted with a polynomial. Note that an ellipsoid fit, the standard for measuring contact angles on flat surfaces would produce inaccurate results as pinning can create three-dimensional variations in the interfacial shape.
For each parameter combination, ten experiments were conducted and averaged. The scatter between the experiments was treated as experimental uncertainty.
The same method was used to determine the contact angle on a flat portion (outside of the structured section of the surface $\theta_{FS}$, which serves as a rough approximation of Young's contact angle, which was measured to be $\theta_{FS}=75^{\circ}\pm 5^{\circ}$ on untreated PMMA and $\theta_{FS}=120^{\circ}\pm 5^{\circ}$ on polymerized PMMA, which is consistent with the measurements and predictions of Ribeiro et al.~\cite{Ribeiro2021}.

While for the pyramids and staggered cubes, symmetry was assumed and the average contact angle from both sides was recorded in the SCA20 program, the ramps could feature different contact angles on the left and right sides. 
%An example image illustrating such a case is depicted in figure~\ref{fig:rampasymmetry}. 
Thus, for the ramps, the two sides were treated separately and in the evaluation, the left side was shifted by an azimuthal angle of $180^{\circ}$ for averaging.

\begin{figure}[!ht]
    \footnotesize
    \includegraphics{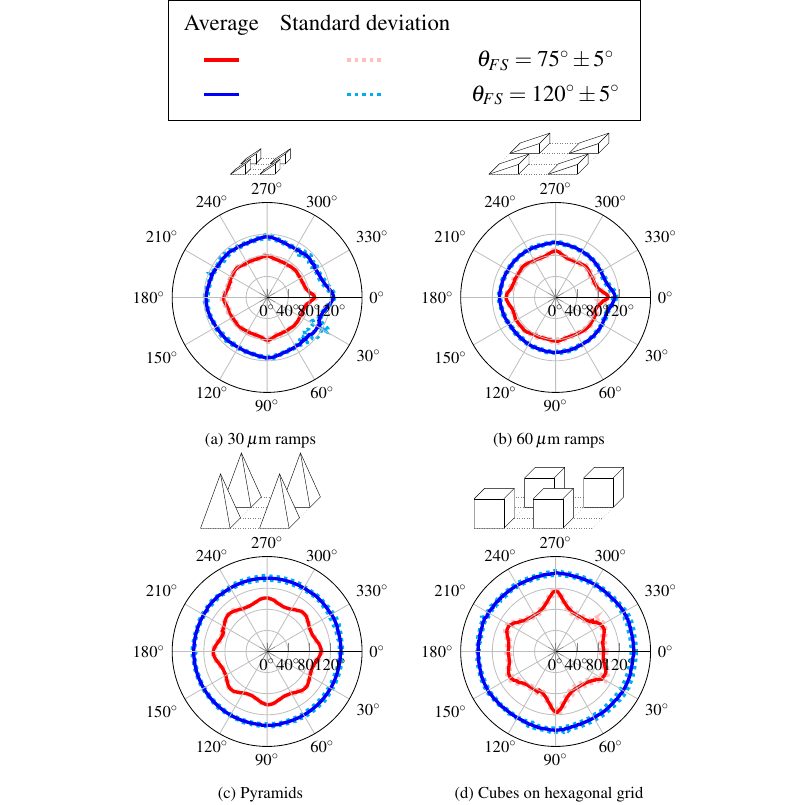}
    \caption{360$^{\circ}$ static apparent contact angle measurements (average and standard deviation between 10 repetitions) of water on untreated and polymerized PMMA surfaces. Azimuthal variations are reduced by polymerization, but the structure still affects the apparent contact angle significantly.}
    \label{fig:360CA}
\end{figure}

%\subsection{Image processing for 360$^{\circ}$ static contact angle measurements}

%\subsection{Tomographic reconstruction of the droplets}

%\section{Evaluation of the surfaces and their wetting behavior}

\section{360$^{\circ}$ static contact angle measurements}
Figure~\ref{fig:360CA} presents the apparent contact angle variation over all azimuthal angles for water on the new microstructured surfaces. Both untreated ($\theta_{FS}=75^{\circ}\pm 5^{\circ}$~\cite{Ribeiro2021}, marked in red) and polymerized surfaces ($\theta_{FS}=120^{\circ}\pm5^{\circ}$~\cite{Ribeiro2021}, marked in blue) are shown.
Some interesting features can be observed in the azimuthal variation of the apparent contact angles:
For the staggered cubes, symmetry is observed along two separate axes and the highest contact angles are reached along the axial direction with unobstructed grooves. The second highest contact angles can be found along the secondary axes of the hexagonal grid which are located at $60^{\circ}$ and $120^{\circ}$. Here, however, the contact angles are somewhat smaller than along the axis.

For the ramps the number of symmetry axes is reduced. 
At the sides of the ramps, smaller peaks are visible, corresponding to a smaller contact angle increase orthogonal to the ramp slope direction than in the directions of the ramp slope.
Lastly, on the "low side" of the ramps, a peak smaller than that on the "high side" of the ramps is visible. Here, the contact line is observed to stick on the inclined surface of the ramp.

\begin{figure}[!ht]
    \centering
    \includegraphics{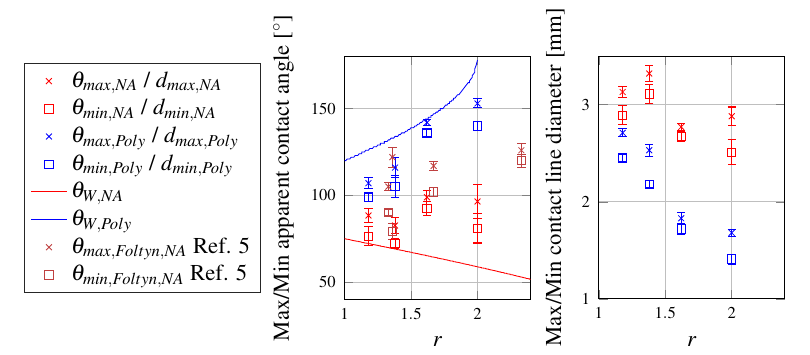}
    \caption{Maximum (x) and minimum (square) contact angles and contact line diameters over the roughness factor for the present experiments both untreated (NA) and polymerized (Poly) and those of Foltyn et al.\cite{Foltyn2021a} (untreated). Solid lines represent theoretically predicted contact angles using the Wenzel model. Error bars represent the standard deviations between experimental repetitions also given in table~\ref{tab:avgCA}.}
    \label{fig:maxminangles}
\end{figure}

\begin{figure}[!ht]
    \footnotesize
    \includegraphics{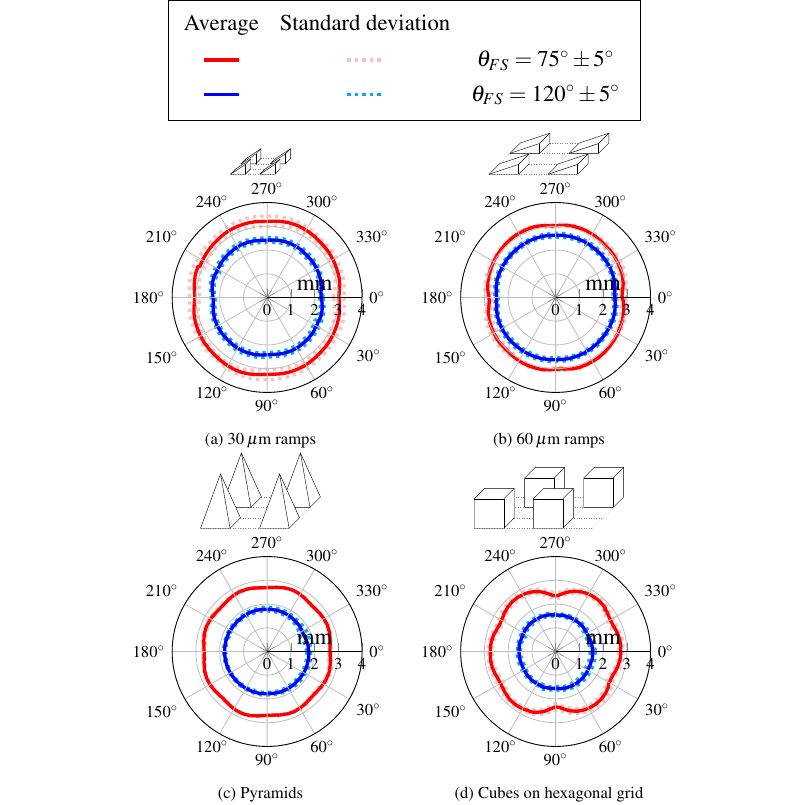}
    \caption{Contact line diameters at azimuthal angles showing elongation of the diameter orthogonally to the ramp directions  (average and standard deviation between 10 repetitions), a rounded octagonal shape on the pyramids and strong pinning effects on the cubes on the hexagonal grid.}
    \label{fig:360Dia}
\end{figure}

\begin{figure}[!ht]
    \centering
    \includegraphics{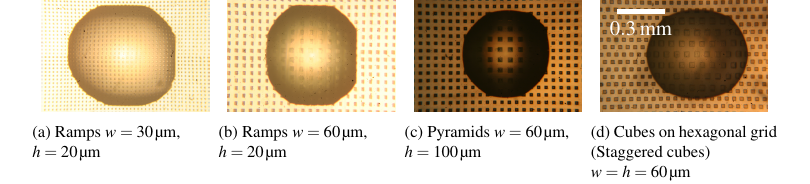}
    \caption{Top view light microscope images of sessile droplets on untreated structures $\theta_{FS} \approx 75^{\circ}$. Other than the contact line diameters, the projected view of the droplet on the staggered cubes shows a rounded elongated hexagon.}
    \label{fig:topviewstruct}
\end{figure}

For polymerized surfaces, the azimuthal variation of contact angles is generally smaller than for untreated surfaces and the ramps again feature a contact angle closer to the smooth-surface value of $\theta_{FS}\approx 120^{\circ}$. For the pyramids and staggered cuboids, a significant increase of the apparent contact angle is observed to almost $150^{\circ}$, indicating that in this case, the structures make the hydrophobized surfaces even more hydrophobic which is consistent with both the Cassie-Baxter and Wenzel model. An overview of the maximum and minimum contact angles over all azimuthal angles is shown in figure~\ref{fig:maxminangles}. The Wenzel model can be seen to rarely correspond to either the maximum or minimum contact angles, but those observed are generally less hydrophobic/hydrophilic than the model. For comparison, the data of Foltyn et al.\cite{Foltyn2021b} for untreated cuboids are also displayed. For their structures and the structures investigated here, no clear trend in the contact angles can be observed for the untreated surfaces in dependence of the roughness factor $r$. However, for the polymerized surfaces, such a trend is apparent, as the contact angle rises with increasing roughness factor.

The anisotropic behavior of the contact angles can be compared to the azimuthal variations of the contact line diameter shown in figure~\ref{fig:360Dia}. Most notably, a local increase in the contact angles normally corresponds to a decrease in the contact diameter, creating straight apparent contact lines. This applies to all surface types and inherent wettabilities. Again, the difference between the polymerized and untreated state appears largest for the staggered cubes and pyramids, while the ramps only feature minor changes, with a larger diameter on the diagonals and the lowest diameter in the ramp-slope directions. The azimuthal variations of the contact line are also much smaller (both absolute and relative) for the polymerized surfaces than for the untreated cases. Note that the contact diameter is always based on the center between the two visible three-phase contact points. As such, the evaluation of the contact diameter is symmetric and differences between two diametrically opposite sides are not resolved. 

For untreated pyramids, an octagon-like shape is observed while for staggered cubes, a petal-like shape can be seen. The octagon-like shape of the contact line is confirmed using top-view light-microscope images of a sessile droplet as seen in figure~\ref{fig:topviewstruct}. Similarly, the petal-shape for the staggered cubes in the contact line diameter appears as a rounded elongated hexagon in the microscope image. Likely, the strong variations in the contact angle can cause the visual obstruction of the local contact line in the side view. As such, the contact line seen in figure~\ref{fig:360Dia} is "smoothed". For both ramped structures, a straight pinned line is observed in figure~\ref{fig:topviewstruct} on the high side and lateral side directions of the ramps while the ramp-down direction shows a gentle curve. Consistently with the contact line diameters measured from the side view, the extent of the spread droplet is larger along the axes orthogonal to the ramp axes.

Similarly to the contact angle evaluation, maximum and minimum contact line diameters are shown in figure~\ref{fig:maxminangles}. For untreated surfaces, again, no clear dependence is found, but for polymerized surfaces, both the maximum and minimum contact line diameters reduce with increasing roughness factor $r$, which is consistent with the Wenzel and Cassie-Baxter models.

It should be noted that the measured static contact angles will be closer to advancing contact angles than receding contact angles, as the measurements were performed after deposition and a subsequent "settling" wait time. Attempts to measure the receding contact angles were made, but for untreated surfaces ranged at near-zero values where the measurement methods were inaccurate. However, this indicates that for untreated surfaces, pinning occurs also during the receding process. Thus, a full wetting Wenzel state is assumed for all untreated surfaces. For the polymerized surfaces, the receding contact angles were similar to the measured static contact angles. Here, it is not entirely clear from the data discussed here if the achieved wetting state is a Wenzel or Cassie-Baxter state.

\section{Contact line pinning at surface features}

%Similarly to what was observed by Foltyn~\cite{FoltynPhD}, the observed contact angles for both untreated and polymerized surfaces are larger than for the respective flat surface samples. This contradicts predictions of Wenzel~\cite{Wenzel1936}, whose equation suggests that the apparent contact angle is reduced for the untreated surfaces.

Apart from the azimuthal variation, Foltyn et al.~\cite{Foltyn2021b} observed an increase in the maximum and even the minimum apparent contact angles over the flat surface contact angle.
Similarly, in the present study, neither the maximum, minimum nor average contact angles match those observed on a flat surface.
For a comparison with the simple theories discussed, Wenzel and Cassie-Baxter contact angles for the given surfaces structures at the two inherent wettabilities are given in table~\ref{tab:WettabilityOverview}. Note that for most combinations, equation~\ref{eq:WCBtrans} predicts that the Wenzel state is stable. Thus, the Wenzel contact angles are given in the table for most surface samples with the exception of polymerized staggered cubes, for which the Cassie-Baxter contact angle is provided. Here, the critical contact angle is smaller than the inherent contact angle of a polymerized surface ($\theta_{FS} = 120^{\circ}$). When comparing the theoretical contact angles with the predicted contact angles, a qualitative correspondence to theory is observed in the experiments for polymerized surfaces, while for untreated surfaces no such correlation is found in this study. This means that more complex models are necessary to determine the wetting behavior of such microstructured surfaces.

\vspace{10pt}
\begin{table}[!htb]
\small
\centering
	\begin{tabular}{|l|c|c|c|c|c|}
		\hline 
		Treatment/Structure 			& Flat 		& \qty{30}{\micro\meter} ramps & \qty{60}{\micro\meter} ramps & Pyramids & Staggered Cubes\\ \hline 
		None					& $\theta_{FS}\approx75^{\circ}$ & $\theta_W\approx69^{\circ}$ & $\theta_W\approx72^{\circ}$ & $\theta_W\approx65^{\circ}$ & $\theta_W\approx58^{\circ}$ \\ \hline
		Polymerized				& $\theta_{FS}\approx120^{\circ}$ & $\theta_W\approx133^{\circ}$ & $\theta_W\approx126^{\circ}$ & $\theta_W\approx144^{\circ}$ &	$\theta_{CB}\approx151^{\circ}$ \\ \hline
		Crit Young angle				& - & $\theta_{Y,crit}\approx131^{\circ}$ & $\theta_{Y,crit}\approx143^{\circ}$ & $\theta_{Y,crit}\approx123^{\circ}$ &	$\theta_{Y,crit}\approx115^{\circ}$ \\ \hline
	\end{tabular}
	\caption{Inherent and theroretical Wenzel/Cassie-Baxter contact angles of PMMA samples for different treatments}
	\label{tab:WettabilityOverview}
\end{table}

For the ramps, the highest peak is observed on the "high" side of the ramps. Using the images and contact angle measurements it is apparent that the contact lines are pinned at the ramp tops, causing a local increase in the apparent contact angle. Here, the canthotaxis angle is maximized, as the inherent structure angle is larger than $90^{\circ}$.
Pinning at the pillar tops was previously also reported by Foltyn et al.~\cite{Foltyn2021b} for cuboid pillars. As the structures are regular, the canthotaxis\cite{Langbein2002} effect at the sharp edges can determine the maximum apparent contact angle as the droplet advances. The canthotaxis effect means that further spreading is only possible if the local contact angle surpasses the local inherent contact angle plus the surface feature angle. As the surface feature angle is highest at the pillar tops ($>90^{\circ}$) the canthotaxis effect is strongest in this location. At the sides, the canthotaxis effect likely also plays a role, but has a smaller effect as the fluid can circumvent the large angle at the low side of the ramp. For the contact angle peak at the low side of the ramps, figure~\ref{fig:topviewstruct} suggests that the contact line is located somewhere on the slope of the ramps. This increases the apparent contact angle here to a value between the contact angle on flat PMMA $\theta_{FS}$ and the inherent contact angle plus the slope of the ramp $\theta_{FS}+\arctan{\left(h_s/w_s\right)}$.  Chu et al.\cite{Chu2010} reported similar results for bent rods on a surface while Song et al.\cite{Song2018} observed such behavior for nanometric droplets on nanometric structures.

For the pyramids and staggered cubes, the preferred axes for the contact lines between two surface elements likely determine the apparent contact angles. Interestingly, for the pyramids, the diagonals appear to be a second dominant pinning line apart from the structure axes and behaves comparably to the structure axis. 
The anisotropy in the staggered cubes is likely comparable to that observed in the grooved surfaces investigated by Foltyn et al.\cite{Foltyn2021b}. In their work, the preferred pinning axis corresponds to the groove directions, which is comparable to the primary groove axis in the staggered cubes (the horizontal axis in figure~\ref{fig:topviewstruct} (d)).

\section{Conclusions and Outlook}
In this study, the effect of microstructured surfaces on the azimuthal variation of the static contact angle of sessile water droplets was reported. This provides new insights into the wetting behavior on regularly rough and structured surfaces and could provide a starting point for the design of surfaces to enhance directional liquid transport~\cite{Yada2021,Yada2022}. Other than in previous studies which mainly used simpler linear grooves, cuboids or cylinders, we used staggered cubes, ramps with two different characteristic sizes and pyramids at two different inherent wettabilities. 
Notably, on polymerized surfaces, a smaller azimuthal variation of the contact angles is observed than on untreated surfaces and the reported contact angles are much closer to the theoretical predictions of using the Wenzel\cite{Wenzel1936} and Cassie-Baxter\cite{Cassie1948} contact angles. Discrepancies between theoretical predictions and experimental results were observed for untreated surfaces which likely result from contact line pinning. This is corroborated by an analysis of the contact line diameters and backlight images from the top of sessile droplets.

In the future, the dynamic wetting behavior, including advancing and receding contact angles could be studied for the present surfaces. Additionally, their effect on dynamic droplet impacts could be investigated.

\section*{Appendix: Tabulated average, maximum and minimum contact angles and diameters}

\begin{table}[h]
    \centering
    \begin{tabular}{l|c|c|c|c}
        Structure & \qty{30}{\micro\meter} ramps & \qty{60}{\micro\meter} ramps & Pyramids & Staggered cubes \\
        \hline
        $\theta_{avg,untreated}$ & $76.1^{\circ}\pm 4.3^{\circ}$ & $83.2^{\circ}\pm 5.2^{\circ}$ & $95.4^{\circ}\pm 4.4^{\circ}$ & $91.0^{\circ}\pm 4.0^{\circ}$  \\
        $\theta_{avg,Poly}$  & $111.0^{\circ}\pm 5.5^{\circ}$ & $103.0^{\circ}\pm 3.3^{\circ}$ & $139.3^{\circ}\pm 3.1^{\circ}$ & $146.9^{\circ}\pm 4.0^{\circ}$\\
        \hline
        $d_{avg,untreated}$ / mm & $3.25\pm 0.11$ & $3.02\pm0.12$ & $2.73\pm0.06$ & $2.68\pm0.15$ \\
        $d_{avg,Poly}$ / mm & $2.41\pm 0.11$ & $2.60\pm 0.08$ & $1.78\pm 0.08$ & $1.56 \pm 0.08$ \\
        \hline
        $\theta_{max,untreated}$ & $82.6^{\circ}\pm 4.7^{\circ}$ & $88.4^{\circ}\pm 3.9^{\circ}$ & $98.9^{\circ}\pm 4.0^{\circ}$ & $96.5^{\circ} \pm 9.7^{\circ}$ \\
        $\theta_{max,Poly}$ & $116.4^{\circ}\pm 5.6^{\circ}$ & $107.3^{\circ}\pm 3.1^{\circ}$ & $142.0^{\circ}\pm 2.0^{\circ}$ & $153.1^{\circ} \pm 2.6^{\circ}$ \\
        \hline
        $d_{max,untreated}$ / mm & $3.32\pm0.08$ & $3.13\pm0.06$ & $2.77\pm0.04$ & $2.88\pm 0.10$ \\
        $d_{max,Poly}$ / mm & $2.53\pm 0.06$ & $2.71\pm0.04$ & $1.83\pm0.06$ & $1.68\pm 0.04$ \\
        \hline
        $\theta_{min,untreated}$ & $72.2^{\circ}\pm2.7^{\circ}$ & $76.4^{\circ}\pm5.6^{\circ}$ & $92.4^{\circ}\pm4.0^{\circ}$ & $81.0^{\circ}\pm8.5^{\circ}$ \\
        $\theta_{min,Poly}$ & $105.1^{\circ}\pm6.2^{\circ}$ & $99.0^{\circ}\pm2.4^{\circ}$ & $136.1^{\circ}\pm1.9^{\circ}$ & $140.2^{\circ}\pm2.8^{\circ}$ \\
        \hline
        $d_{min,untreated}$ / mm & $3.11\pm0.10$ & $2.89\pm0.10$ & $2.67\pm0.05$ & $2.51\pm 0.13 $ \\
        $d_{min,Poly}$ / mm & $2.18\pm0.04$ & $2.45\pm0.04$ & $1.72\pm0.05$ & $1.41\pm 0.05$ \\
    \end{tabular}
    \caption{Average minimum and maximum contact angles and contact line diameters for all structures, polymerized and untreated.}
    \label{tab:avgCA}
\end{table}

\section*{Acknowledgments}
We appreciate the financial support of this work by the Deutsche Forschungsgemeinschaft	(DFG) in the framework of the International Research Training Group “Droplet Interaction Technologies” (GRK	2160/2: DROPIT) under project number 270852890. We further thank the Karlsruhe Nano Micro Facility, a Helmholtz Research Infrastructure at Karlsruhe Institute of Technology at which the surfaces were manufactured under KNMF Proposal 2021-026-030550.
We also acknowledge the contributions in the manufacturing and characterization of the surfaces of Samuel Bergdolt (fabrication), Marc Schneider (replication), Marco Ehrhardt (electroplating) and Markus Wissmann (characterization).

\section*{Data Availability Statement}
The data that support the findings of this study are available from the corresponding author upon reasonable request.
	
\section*{Author contributions statement}
P.P. conceptualized the study, performed the apparent contact angle measurements, analysed and interpreted the results and wrote the main manuscript text. S.H. and M.G., manufactured the micro-structured surfaces and co-wrote the manuscript text. S.H. fabricated and validated the surfaces, reviewed and edited the main manuscript text and administered the manufacturing project.
M.G. performed the electroforming and chemical treatment steps of the manufacturing process.
M.W. reviewed and edited the manuscript.
B.W. conceptualized the study, interpreted the results, administered the project, acquired funding and edited/reviewed the manuscript. All authors have read and agreed to the published version of the manuscript.

\section*{Competing interests}
The authors declare no competing interests.

\section*{Additional information}

Correspondence and requests for material should be addressed to P.P.

%\section{Appendixes}

\section*{Data Availability Statement}
Data supporting the findings of this study are available from the corresponding author upon request.
\bibliography{PPLib}% Produces the bibliography via BibTeX.

\end{document}